\def\year{2022}\relax
\documentclass[letterpaper]{article} % DO NOT CHANGE THIS
\usepackage{aaai21}  % DO NOT CHANGE THIS
\usepackage{times}  % DO NOT CHANGE THIS
\usepackage{helvet} % DO NOT CHANGE THIS
\usepackage{courier}  % DO NOT CHANGE THIS
\usepackage[hyphens]{url}  % DO NOT CHANGE THIS
\usepackage{graphicx} % DO NOT CHANGE THIS
\usepackage{natbib}  % DO NOT CHANGE THIS AND DO NOT ADD ANY OPTIONS TO IT
\usepackage{caption} % DO NOT CHANGE THIS AND DO NOT ADD ANY OPTIONS TO IT
\usepackage{xcolor}

\newcommand{\galen}[1]{}
\newcommand{\tim}[1]{}
\newcommand{\amy}[1]{}
\newcommand{\new}[1]{#1}

\newcommand{\appendixlink}[0]{Appendix}

\usepackage{xspace}
\newcommand{\hide}[1]{}
\newcommand{\xhdr}[1]{\vspace{1.7mm}\noindent{{\bf #1.}}} % can modify to adjust space -galen

\newcommand{\eg}{\textit{e.g.,}\xspace}
\newcommand{\ie}{\textit{i.e.,}\xspace}
\newcommand{\sect}{\S}

\usepackage{caption}
\usepackage{subcaption}

\usepackage[hidelinks]{hyperref} % hidelinks hides the ugly boxes

\usepackage{multirow}
\usepackage{tabularx}

\title{What Makes Online Communities `Better’? Measuring Values, \\ Consensus, and Conflict across Thousands of Subreddits}

\author{
Galen Weld, Amy X. Zhang, Tim Althoff \\
\normalsize{{\normalfont Paul G. Allen School of Computer Science \& Engineering, University of Washington}} \\
\normalsize{\normalfont \{gweld, axz, althoff\}@cs.washington.edu }
}

\begin{document}

\maketitle
\begin{abstract}
Making online social communities ‘better’ is a challenging undertaking, as online communities are extraordinarily varied in their size, topical focus, and governance. As such, what is valued by one community may not be valued by another.
\new{However, community values are challenging to measure as they are rarely explicitly stated.}
In this work, we measure community values through \new{the first large-scale  survey of community values, including} 2,769 reddit users in 2,151 unique subreddits. Through a combination of survey responses and a quantitative analysis of public reddit data, we characterize how these values vary within and across communities.

\new{Amongst other findings,} we \new{show} that community members disagree about how safe their communities are, that longstanding communities place 30.1\% more \new{importance} on trustworthiness than newer communities, and that community moderators want their communities to be 56.7\% less democratic than non-moderator community members.
\new{These findings have important implications, including suggesting that care must be taken to protect vulnerable community members, and that participatory governance strategies may be difficult to implement.}
Accurate and scalable modeling of community values enables research and governance which is tuned to each community's different values. To this end, we demonstrate that a small number of automatically quantifiable features capture a significant yet limited amount of the variation in values between communities with a ROC AUC of 0.667 on a binary classification task.
However, substantial variation remains, and modeling community values remains an important topic for future work.
We make our models and data public to inform community design and governance.
\end{abstract}

\section{Introduction}\label{sec:intro}

% removed from 'support' to save space de2014mental,althoff2015donor, althoff2014ask,de2016discovering -g
% removed from 'misinfo' to sace space karlova_social_2013, jin_misinformation_2014

Online social communities are extraordinarily varied and capture almost every aspect of our society. Every day, people use millions of online communities to get the news~\cite{weld_political_2021, geiger_2019_online_news, volkova_separating_2017}, for support~\cite{sharma_engagement_2020, Sharma2020ACA, wadden2021effect}, for entertainment~\cite{chen_2021_memes, centivany_popcorn_2016}, to discuss with others~\cite{tan2016winning,zhang2018characterizing,chang2020don}, to compete with others~\cite{shameli2017gamification,althoff2017online}, and many other purposes. Some aspects of communities have been tied to specific societal harms, including distribution of misinformation~\cite{jahanbakhsh_exploring_2021, Tran2020AnIO, anagnostopoulos_viral_2014, zollo_misinformation_2018}, harassment and bullying~\cite{Lenhart2016OnlineHD, jhaver_2018_blocklists, matias_preventing_2019, pater_characterizations_2016, burke_winkelman_exploring_2015, ybarra_how_2008, van_laer_means_2014, bretschneider_detecting_2014, matias_study_2020}, and increasing polarization ~\cite{bessi_social_2016, shao_anatomy_2018, howard_social_2018, grinberg_fake_2019, bossetta_digital_2018, bovet_influence_2019}. 

\new{
However, there are many important aspects of community health beyond these harms, such as the quality of content and the diversity of the community.
Given the immense diversity of online communities, it follows that there is no `one size fits all' approach to making communities `better'~\cite{weld_2021_taxonomy}.
What is strongly valued by members of one community may not be valued by another, and furthermore, members within a community may disagree with one another about what values are most important.
}

It is challenging to measure community values across many communities, as they are infrequently formalized or explicitly enumerated. Some work has attempted to study values implicitly by examining communities' rules~\cite{fiesler_reddit_nodate} or removed content~\cite{chandrasekharan_internets_2018}; however, these approaches only capture values as implemented by moderators~\cite{matias_civic_2019}, and are unable to measure the degree to which community members disagree. 
% \new{To date, no large-scale survey of community members' values has been conducted.}

In this work, we contribute the first large-scale survey of community members' values to date.
Specifically, we survey, analyze, and model community values on reddit.
Using a taxonomy of nine different values (\sect\ref{sec:summary_of_values}) previously developed from qualitative user studies~\cite{weld_2021_taxonomy}, we ask community members about (1) which of these values are most and least important to their community, (2) the current state of each value in the community, and (3) how they would like the community to change with regards to each value (\sect\ref{sec:instrument}).
We recruit survey respondents from a diverse set of reddit users, ranging from very new reddit users to \new{moderators with 10 years of experience.} 2,769 members of 2,151 different subreddits completed our survey, making this survey an order of magnitude larger than previous small-scale surveys~\cite{weld_2021_taxonomy}.

With our participants' consent, we gather their reddit post and comment history, along with metadata and six months of content from each subreddit in our dataset (\sect\ref{sec:user_sub_feats}). Using these data, we answer four research questions:
\begin{itemize} % fixing the labels jumping into the margins
    \setlength{\itemindent}{6mm}
    \setlength{\itemsep}{0em}
    \item [\textbf{RQ1}] What are communities’ values, and how do they vary across communities? (\sect\ref{sec:values_across_communities})
    \item [\textbf{RQ2}] Within communities, where is there disagreement over values? (\sect\ref{sec:consensus})
    \item [\textbf{RQ3}] How do moderators differ in their values from non-moderator community members? (\sect\ref{sec:moderators})
    \item [\textbf{RQ4}] To what degree can community values be predicted based on automatically measurable features? (\sect\ref{sec:prediction})
\end{itemize}

\noindent
We find that there is substantial variation in values both within and across communities, especially with regards to safety, for which there is 47.4\% more disagreement within communities than other values.
We leverage theories of group bonds from sociology~\cite{prentice_1994_common_identity_bond, grabowicz_2013_distinguishing, ren_applying_2007} that suggest that communities built around interpersonal connection place greater emphasis on safety, engagement, and inclusion than communities built around shared interests. We find communities for specific groups of people (\eg /r/teenagers) place 1.21 points (out of 8) more importance on inclusion than communities for the sharing of pictures and video (\sect\ref{sec:values_across_communities}). We examine differences between newcomers and senior community members in the context of literature on the challenges of managing community growth and new members~\cite{lin_2017_better_when_smaller, dnm_2013_old_members} and find that, on reddit, new members are more positive in their perception of the current states of their communities than more senior members (\sect\ref{sec:consensus}). Given that governance on reddit is often characterized by divisions between moderators and non-moderators~\cite{matias_civic_2019}, we measure differences in values between moderators and non-moderators. We find that moderators perceive their communities as 14.7\% less democratic, think they should be 56.7\% less democratic, and that democracy is 23.6\% less important, relative to the average non-moderator in each community. This has important implications for the implementation of participatory governance practices in online communities~\cite{zhang_policykit_2020} (\sect\ref{sec:moderators}).

Given the large amount of variation between communities, we suggest that researchers and community leaders consider the specific values and needs of each community when making decisions about how to change those communities. As measuring community values with survey responses is time-consuming and expensive, the ability to accurately model community values with automatically quantifiable features would be of great value. Through a binary classification task which seeks to differentiate between above- and below-average communities, we show that such features are able to predict a substantial amount of the variation between communities' values with a ROC AUC of 0.667 (\sect\ref{sec:prediction}). However, much variation remains challenging to predict, and additional research is needed on modeling and measuring community values.
We make our models and anonymized responses public to support further research\footnote{\scriptsize \url{https://behavioral-data.github.io/reddit_values_surveys_public/}}.
\section{Related Work}\label{sec:related}

\xhdr{Content Moderation, Rules, and Norms}
A community's formal rules can offer insight into that community's values. On reddit, rules have been studied by \citet{fiesler_reddit_nodate}, who produced a taxonomy of 24 different types of rules in use across 1,000 subreddits. These rules are enforced by volunteer moderators~\cite{matias_civic_2019}, and in some cases, content removed by moderators for violating the rules can be recovered and used to characterize community norms~\cite{chandrasekharan_internets_2018}. However, one significant drawback of these approaches is that rules are both set and enforced by moderators, in almost all cases without any input from non-moderator community members~\cite{zhang_policykit_2020}. As such, such analyses may fail to represent the interests of the non-moderator majority of the subreddit. Further evidence for this can be found in studies of user reactions to moderator actions, which find that there is often conflict and disagreement between moderators and non-moderators \cite{srinivasan_2019_content_removal, jhaver_2019_user_reactions}. In contrast, our method of explicitly surveying both moderators and non-moderators enables us to directly measure the differences between these two groups (\sect\ref{sec:moderators}).

\xhdr{Community Governance}
Nearly all social media communities (\eg Subreddits, Facebook Groups, Twitter) adopt a strictly hierarchical governance model, where each community is managed by a small group of privileged moderators (sometimes also called admins) who have the authority to set rules and enforce them~\cite{zhang_policykit_2020, matias_civic_2019}.
%On some platforms, such as reddit and Facebook Groups, there is some modicum of transparency, as the identity of the moderators is public, but any many platforms (\eg Facebook, Twitter, Instagram) the moderation practices are almost entirely opaque.
On reddit, while moderators are beholden to platform administrators~\cite{jhaver_2021_designing}, they typically have wide latitude to set and enforce policies as they see fit, with no requirement for community input. % into decision making processes. 
Social media communities stand in contrast to many peer-production communities such as Wikipedia, which operates primarily on a consensus model~\cite{halfaker_rise_2013}, or StackExchange, which holds formal elections. While some systems to incorporate democracy into reddit have been developed~\cite{zhang_policykit_2020}, such systems have not been widely adopted, and moderators often face conflict and accusations of corruption~\cite{matias_civic_2019}.
In this work, we ask community members about their perceptions of democracy, and examine how moderators' and non-mods' responses differ (\sect\ref{sec:moderators}).

\xhdr{Growing Pains and Internal Conflict}
Community growth and differences between new and senior members are a frequent source of conflict within communities that have been studied on a range of platforms~\cite{robert_e_kraut_building_2012, halfaker_2011_bite_newbies}. Research investigating this tension on reddit has taken either a high-level approach which relies on implicit signals of conflict such as linguistic change and negative sentiment~\cite{dnm_2013_old_members, lin_2017_better_when_smaller}, or a qualitative interview with a small number of participants, focusing only on a single subreddit~\cite{kiene_2016_eternal_september, cho_2021_potential_new_members}. In contrast, our approach allows us to include over 2,000 communities while still gathering granular information about values through explicit survey questions.

\xhdr{Community Bonds and Membership} 
We draw upon the Theory of Common Identity and Common Bonds~\cite{prentice_1994_common_identity_bond}, which suggests that some communities form due to common identities (\ie shared interests) while others form due to common bonds (\ie social relationships). Some previous work has examined this theory in the context of online communities~\cite{ren_applying_2007, grabowicz_2013_distinguishing}; in contrast, our work explicitly surveys community members on their values.

\section{Methods}\label{sec:method}

\subsection{Measuring Community Values}\label{sec:summary_of_values}
Central to this work is the set of nine values around which we design our survey instrument~(\sect\ref{sec:instrument}). The set of values we use is grounded in sociology literature on different dimensions of social relations~\cite{deri_coloring_2018, bao_conversations_2021, choi_ten_2020} and drawn directly from the taxonomy developed by \citet{weld_2021_taxonomy} via iterative categorization of unstructured survey responses from redditors. The complete \citet{weld_2021_taxonomy} taxonomy consists of 29 different values in nine major categories. As it is impractical to ask about 29 different values, we use the nine major categories with minor modifications; we include both Variety of Content and Diversity of People, we break Offensive, Abusive, and Harassing Content or Behaviors into a separate category called Safety, and we drop the Technical Features category as items within this category are outside the scope of control of community members and moderators. As such, the nine values we consider in this work are listed in Table~\ref{tab:values}. 
For each of these values, we ask community members about three dimensions: (1) the overall importance of the value to their community, (2) their perception of the value's current state in their community, and (3) their desire to change their community with regards to the value.

\begin{table}[t]
\small
    \centering
\begin{tabular}{r|l}
\textbf{Quality}             & Quality of the content                                  \\
\textbf{Variety}             & Variety in/of the content                               \\
\textbf{Diversity}           & Diversity of the people                                 \\
\textbf{Trust}               & Trustworthiness of the people and information           \\
\textbf{Engagement}          & Members' engagement with one another                    \\
\textbf{Inclusion}           & Members' inclusion and ability to contribute            \\
\textbf{Size}                & Size of the community                                   \\
\textbf{Democracy}           & Community input into moderator decisions          \\
\textbf{Safety}              & Absence of offensive or harassing behavior                                            
\end{tabular}
    \caption{We leverage the taxonomy of widely-held values on reddit by \citet{weld_2021_taxonomy}, which was developed through user studies and iterative categorization.}
    \label{tab:values}
\end{table}

\subsection{Reddit Background}\label{sec:reddit}
% We focus on reddit in this work. 
Reddit is the fifth most popular social media site in the United States~\cite{statista_social_media_popularity}, and is an ideal platform for researching the values of online communities as reddit is explicitly divided into many thousands of discrete communities, known as `subreddits.' Each subreddit has its own topic, moderators, rules, and community norms.
%Subreddits are commonly prefixed with \texttt{/r/}, \eg /r/science, a large subreddit for the discussion of science-related news.
Within a subreddit, a user may post a link to another website (a linkpost), some text (a selfpost), or may comment on an existing post. 
Almost all content on reddit is publicly available~\cite{baumgartner_pushshift_2020}, and reddit has been widely studied~\cite{medvedev_anatomy_2019}.

\subsection{Data Collection: Online Survey}\label{sec:instrument}
Responses were gathered through an online survey hosted on the Qualtrics platform. We summarize the survey here, a complete copy is online\footnote{\scriptsize \url{https://behavioral-data.github.io/reddit_values_surveys_public/}}. The survey consists of three parts: (1) informed consent, (2) general reddit questions, and (3) subreddit-specific questions. Before any other questions are asked, the participant is shown a brief summary of the survey, study, and IRB information (\sect\ref{sec:ethics}) and asked for their consent. After this point, all questions are optional.

The general reddit questions ask the participant about their usage of the platform across all subreddits. First, the participant is optionally asked to provide their reddit username, which is used to query the reddit API for their post/comment history. Then, the participant is asked about how often and how much time they spend on reddit, how frequently they `lurk' vs. posting or commenting, how often they browse content aggregated from multiple subreddits (\eg on their front page), and their mobile vs. PC usage of reddit. At the end of this section, the participant is asked to select up to three subreddits that they consider themselves a member of. For reddit users who choose to provide their username, subreddits from their recent post and comment history are automatically suggested.

The subreddit-specific section of the survey asks questions specific to the subreddits the participant listed themselves as a member of. For each subreddit, the participant is asked separately about nine different community values (\sect\ref{sec:summary_of_values}). For each value, the participant is asked about their perception of the \textit{current state} of the subreddit on an 11-point rating scale (\eg Safety: `How much offensive or harassing behavior is there in /r/science?' with scale ends `Lots of offensive behavior' and `No offensive behavior') and their \textit{desired change} for the subreddit on a 3-point rating scale (\eg Safety: `Would you change the safety of /r/science?' with options of `The community should be less focused on safety,' `the focus on safety is about right,' or `The community should be more focused on safety.')
Last, the participant is asked to rank all nine values in order from most important to least important to their experience in the subreddit.

The survey was piloted with 13 participants from a variety of departments at two large American universities. All pilot participants reported no difficulties completing the survey. % having no

\xhdr{Participant Recruiting and Incentives}
Survey participants were recruited through multiple channels, including reddit advertisements, private messages, and distribution on /r/SampleSize, a subreddit for the recruiting of survey participants. Community moderators were additionally recruited via reddit moderator mail. Responses were collected from May-July 2021, with a total of 2,769 people participating. Additional details on recruiting, participation, and compensation are included in \appendixlink~\ref{app:recruiting}.

\xhdr{Quantifying Community Values and Disagreement}
We compare values at the subreddit level instead of \new{the} survey response level, in order to avoid biasing our findings towards particularly popular communities that may receive a larger number of responses.
To measure the degree of (dis)agreement on values at the subreddit level, we compute the mean average deviation (MAD) from the subreddit mean by computing the mean difference between each response for a subreddit and the average response for that subreddit.

\xhdr{Ensuring Response Validity}
\new{
Generally only a subset of community members will respond to our survey. 
To ensure reasonably representative results when extrapolating from survey responses, we exclude subreddits with fewer than 15 responses from our analyses. This threshold was selected through an empirical power analysis (\appendixlink~\ref{app:num_responses}) leveraging subreddits with a high number of responses, which indicated that subreddit averages have stabilized at this number of responses. 
}

% moving here to help with flow
\begin{figure*}[t]
    \centering
    \includegraphics[width=\textwidth, trim={32mm 11mm 25mm 10mm}]{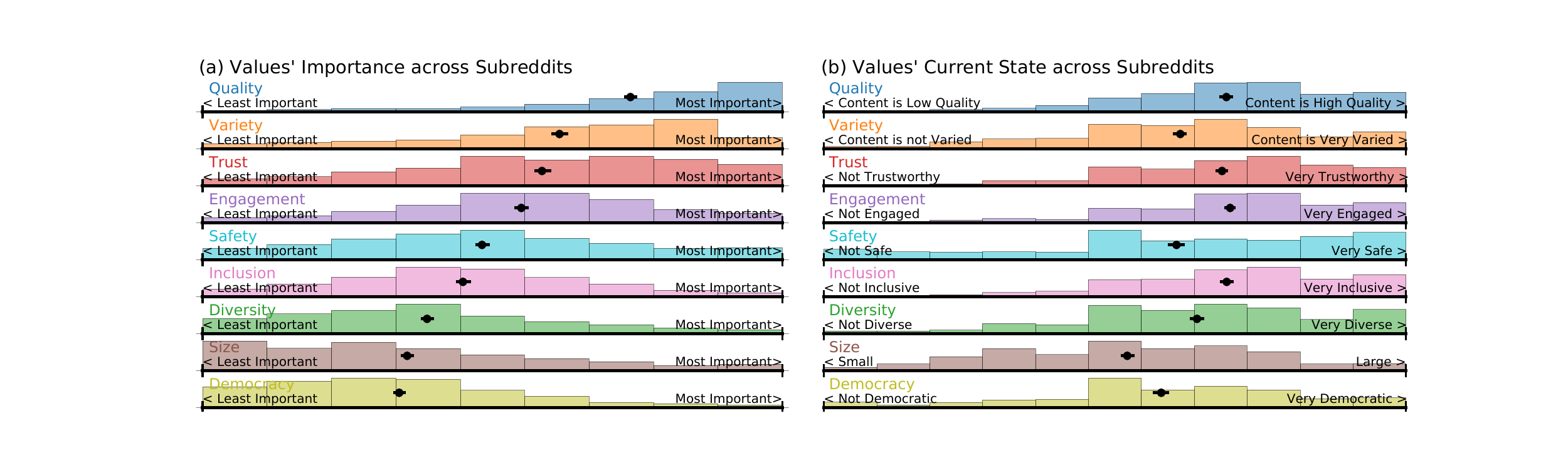}
    \caption{To understand what communities’ values are, we average all responses for each community. (a) shows the distribution of the relative importance of each value across communities. Quality of Content is most frequently considered the most important value, while Size and Democracy are generally considered to be the least important. (b) shows the distribution of communities' perception of their current state. Black points indicate the average community. In this and all figures, bars indicate 95\% bootstrapped confidence intervals.}
    \label{fig:subs_summary}
\end{figure*}

\subsection{Data Collection: User \& Subreddit Information}\label{sec:user_sub_feats}
To augment survey responses from participants, we additionally compute user and subreddit features from publicly available reddit data. We source data from two locations: for each participant in the survey, we download their entire public reddit history and metadata such as account age from the reddit API. To more comprehensively characterize entire subreddits, we extract the most recent six months (January-June 2021) of posts and comments for each subreddit that participants in the survey are members of, using the Pushshift reddit corpus~\cite{baumgartner_pushshift_2020}.

User features are computed from the participant's entire public reddit history, and include the age of their account, their total number of posts, linkposts, selfposts, comments, as well as the mean length (\# of characters) of each of the previous, along with their ratio of posts:comments, ratio of selfposts:posts, the mean and cumulative scores (\# upvotes-downvotes) of their posts and comments, and the mean number of comments received for each of their posts.
Then, for each subreddit the user is a member of, we extract the same set of features while considering only content from that subreddit.
Finally, we compute the fraction of a user's total posts across all of reddit that are in the subreddit(s) they indicated they were a member of. For example, if a person answers survey questions for /r/science, we compute their number of posts in all subreddits, their number of posts in /r/science only, and the fraction of their posts (in any subreddit) which are in /r/science.

Subreddit features are computed from the most recent six months of posts and comments (January-June 2021) in that subreddit. These features include the age of the subreddit, the number of posts, linkposts, selfposts, and comments, as well as the number of removed (by moderators) and deleted (by their author) posts and comments, and the number of distinct users and subscribers each subreddit has. We also compute the mean score of posts and comments in each subreddit, the number of posts/comments per distinct user, and the number of rules declared by the community moderators.

\xhdr{Categorizing Subreddit Topics}
For the 122 largest communities, we additionally hand-label the community topic.\footnote{We additionally experimented with pre-computed subreddit embeddings~\cite{kumar_community_2018, martin_community2vec_2017, waller_generalists_2019}, but these did not explain significant variation in values.} For more details on this taxonomy, see \appendixlink~\ref{app:categories}. The six topic categories we use are:
\textbf{Hobby} communities \eg~/r/nba, /r/bicycling,  
\textbf{Discussion} communities \eg~/r/AskReddit, /r/relationship\_advice, 
\textbf{Media-sharing} communities \eg~/r/pics, /r/CrappyDesign, 
\textbf{News} communities \eg~/r/worldnews, /r/science, 
\textbf{Meme} communities \eg~/r/dankmemes, /r/me\_irl, and
\textbf{Identity-based} communities \eg~/r/india, /r/teenagers.

\subsection{Ethical Considerations}\label{sec:ethics}
\new{We strongly believe that this work will have a positive broader impact by informing the design of online communities in a manner which is aligned with the values of their members.}
\new{The most serious potential negative impact of this work is the potential for deanonymization of responses. We take this possibility seriously and have taken numerous steps to mitigate this risk.}
To ensure the anonymity of our participants, we do not publish their usernames \new{nor any of their reddit usage data,} and remove responses from subreddits whose names or small size could enable deanonymization of individual contributors to our public dataset.
% identify their members from our public dataset. 
All participants were informed of the goals of the study and how we would use and share their data before consenting to participate. In a separate step of the survey, we collect specific additional consent to access users' public reddit histories and use them for research~\cite{Fiesler2018ParticipantPO}, \new{which we do not publish}.
This study was approved by the University of Washington IRB under ID number STUDY00011457.

\section{RQ1: What Are Communities' Values, and How Do They Vary across Communities?}\label{sec:values_across_communities}  
Understanding what communities' values are in general, and how these values vary from community to community, are key questions with implications for community design that also provide context for further analyses in this paper. In this section, we begin by quantifying what values are most important to communities, the current state of these values, and the level of variability across communities. Then, we explore how these values vary across communities according to community topic, age, and size of community.
Informed by Common Identity and Common Bond Theory (\sect\ref{sec:related}), we hypothesize that communities with relatively strong interpersonal relationships, such as Identity-based communities, smaller communities, and older communities, will place greater emphasis on values related to interaction with community members, such as Inclusion, Engagement, and Safety. On the other hand, we hypothesize that larger, younger, and more content-consumption focused communities based on shared interests will place greater emphasis on Quality and Variety of Content and Size.

\xhdr{Method}
To analyze how values vary across communities, we group communities based on their topical category (see \appendixlink~\ref{app:categories} for details on categorization methodology) or by their quartile along a variable of interest, and then average across all communities in each group. When appropriate, we make minor adjustments from true quartile values to improve interpretability. We operationalize community age and size by the time since the community was founded and its number of unique contributors from the reddit API, respectively. We operationalize the degree to which the community is text-based by computing the fraction of text-posts (called selfposts on reddit).

\xhdr{Results}
We find that there is substantial variation in both the importance and current state of values from community to community (Fig.~\ref{fig:subs_summary}). On average, Quality of Content is the most important value, with Size and Democracy generally considered the least important (Fig.~\ref{fig:subs_summary}a).
Safety is especially varied with regards to both its importance (Fig.~\ref{fig:subs_summary}a) and current state (Fig.~\ref{fig:subs_summary}b), with a standard deviation  7.0\% and 20.07\% larger than those of any other values' importance and current state, respectively.
While the average community rates Safety 5/9 in terms of importance, 171 communities have Safety as their most important value, and 176 have Safety as their least important value.

\begin{figure}
    \centering
    %trim=left bottom right top
    \includegraphics[width=\linewidth, trim={0 5mm 0 5mm}]{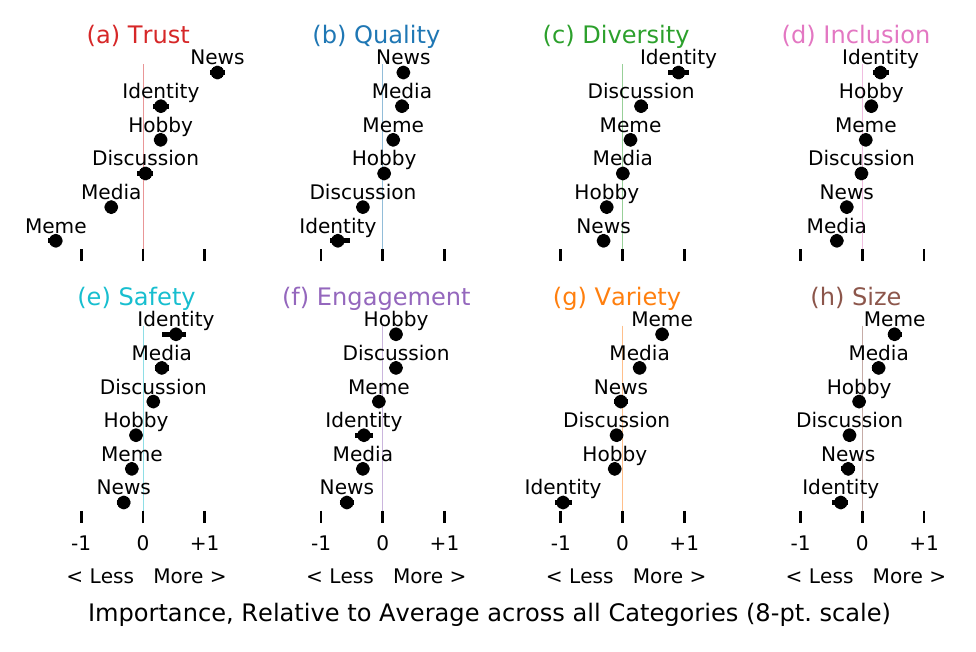}
    \caption{Differences in value importance across communities of different topics. News Communities rate Trust as 2.62 points more important than Meme Communities (out of an 8 point scale). Diversity and Inclusion are especially important to Identity-based Communities. Variety of Content and Size are especially important to Meme and Media-sharing Communities.}
    \label{fig:community_categories}
\end{figure}

Our hypothesis that communities with strong interpersonal relationships will place greater emphasis on community-focused values such as Inclusion, Engagement, and Safety is largely upheld by our results.
Identity-based Communities place greater than average importance on Diversity and Inclusion (Fig.~\ref{fig:community_categories}c,d), while Hobby and News Communities place greater importance on Quality (Fig.~\ref{fig:community_categories}b). Meme and Media-sharing Communities both place higher than average importance on Variety of Content and Size, which includes the amount of content submitted (Fig~\ref{fig:community_categories}g,h).
Identity-based Communities rate Inclusion as 1.21 points (out of 8) more important than Media Communities (Fig.~\ref{fig:community_categories}d).
It is important to note that Common Identity and Common Bond Theory~\cite{prentice_1994_common_identity_bond} does not explain all observed differences between community categories.
News and Meme Communities are both primarily motivated by shared interests, yet News Communities rate Trust as 2.62 (out of 8) points more important than Meme Communities (Fig.~\ref{fig:community_categories}a).

% moving here to help with flow

\begin{figure}
    \centering
    \vspace{4mm}
    \includegraphics[width=\linewidth, trim={0 5mm 0 7mm}]{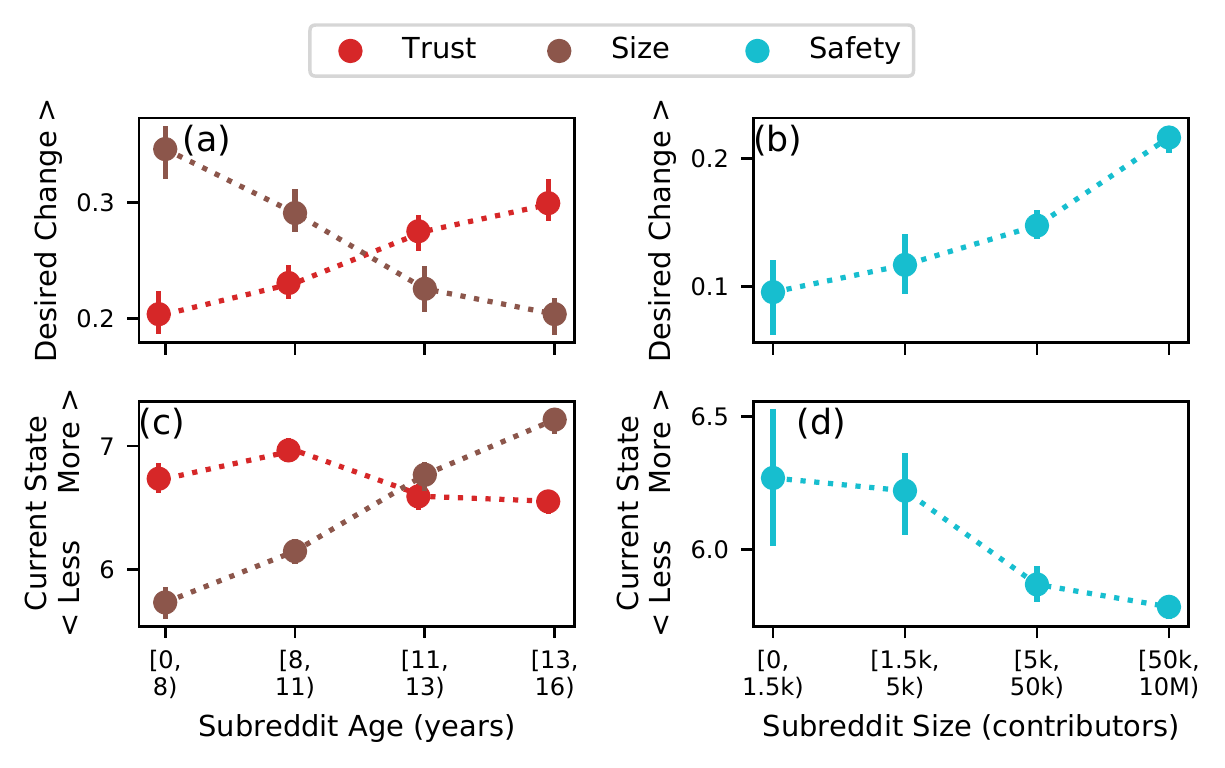}
    \caption{Average importance and desired change across community, binned into approximate quartiles by the age (since founding) and size.  (a) Older communities 30.1\% more strongly desire increased Trust than younger communities. (b) Larger communities have a 126.6\% stronger desire to improve Safety than the smallest communities.}
    \label{fig:sub_age_size}
\end{figure}

When examining the differences between new and older communities, and between small and large communities, differences are especially pronounced for Trust, Size, and Safety (Fig.~\ref{fig:sub_age_size}). The youngest quartile of communities (established within the past 8 years) have a 41.2\% (0.35 vs 0.24) stronger desire to grow than older communities, while older communities have a 30.1\% (0.27 vs. 0.20) stronger desire to improve Trust than younger communities (Fig.~\ref{fig:sub_age_size}a), which is consistent with our hypothesis that older communities are more focused on common bonds than younger communities. Interestingly, this stronger desire to build Trust  holds despite a lack of large difference in the perceived current state of Trust across older and younger communities (Fig.~\ref{fig:sub_age_size}c). 
However, when examining community size, we find large communities with more than 50,000 contributors have a 126.6\% (0.22 vs. 0.10) stronger desire to improve Safety than the smallest communities with less than 1,500 contributors (Fig.~\ref{fig:sub_age_size}b), in contradiction of our hypothesis that smaller communities would value Safety more due to stronger interpersonal relations in smaller communities.
Another potential explanation is that larger communities have poorer current Safety, as we find a 7.73\% (6.27 vs. 5.78) decrease in perceived Safety amongst larger communities (Fig.~\ref{fig:sub_age_size}d).

\xhdr{Implications}
Different values are dramatically more or less important to different communities, which has profound implications, underlining that there is no `one size fits all' approach to improving online communities~\cite{weld_2021_taxonomy}.
The relatively low importance placed upon Democracy may present challenges for the widespread adoption of systems that seek to implement participatory governance practices in online communities~\cite{zhang_policykit_2020, Kelty2017TooMD}.
We examine Democracy and governance further in \sect\ref{sec:moderators}.
Our finding that some communities consider themselves fairly safe while others consider themselves to be very unsafe (Fig.~\ref{fig:subs_summary}) is consistent with previous findings that toxic behavior on reddit is extremely concentrated in a small number of subreddits~\cite{weld_political_2021}, this could further support the practice of community level moderation interventions~\cite{jhaver_does_2019, chandrasekharan_2017_you, habib_act_2019, shen_discourse_2019, chandrasekharan2021quarantined}. However, it is important to not only exclusively consider Safety by averaging over the values of all members of a community. Vulnerable minorities have an important perspective~\cite{Guinier1994TheTO}, yet inherently members of minority groups are too few to significantly influence the community average. We examine this further in \sect\ref{sec:consensus}.
Finally, our results emphasize the importance of Common Identity and Common Bond Theory (\sect\ref{sec:related}), which can guide researchers in their future work on this topic.

\section{RQ2: Within Communities, Where Is There Disagreement over Values?}\label{sec:consensus}
\begin{figure}
    \centering
    %  trim={<left> <lower> <right> <upper>}
    \includegraphics[width=\linewidth, trim={0 5mm 0 2mm}]{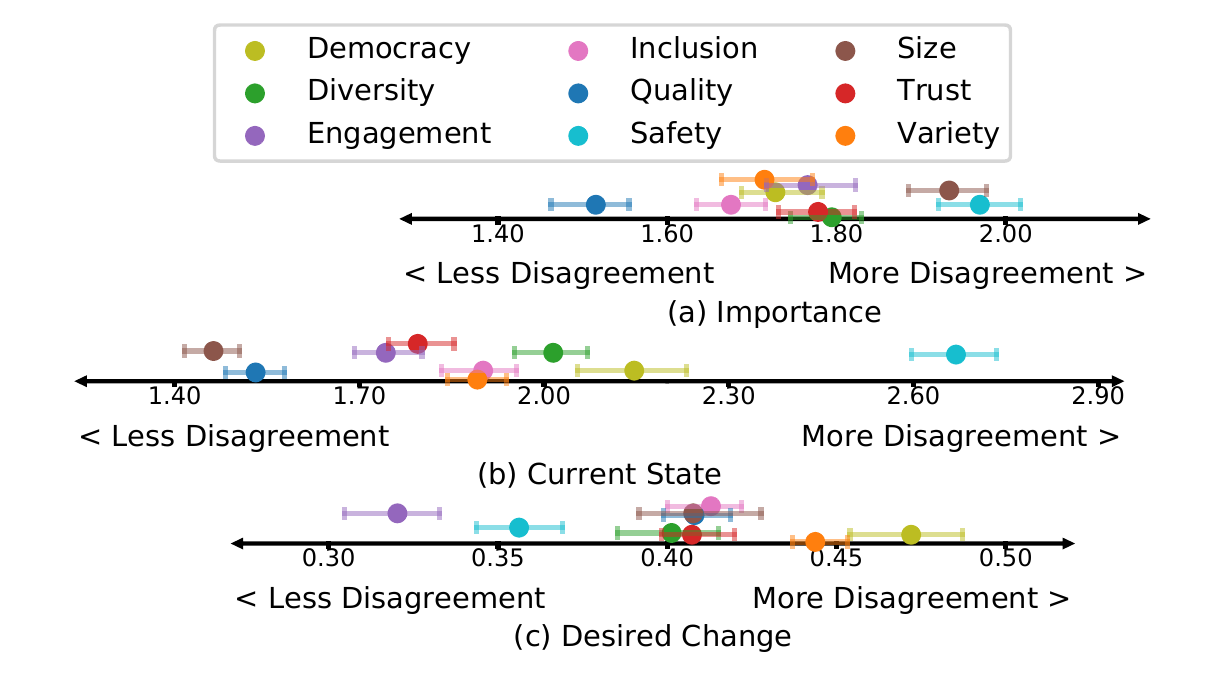}
    \caption{Average disagreement (measured with MAD) in perceptions of importance (a),  current state (b) and desired change (c) across communities. Axes are adjusted for the widths of their respective scales, indicating greater disagreement over the importance of values than their current state and desired change. There is 13.3\% and 47.4\% more disagreement over the importance and current degree of Safety (light blue), respectively, relative to all other values, yet relative consensus on the desire to change Safety.}
    \label{fig:disagreement}
\end{figure}

Understanding where there is consensus on values, and where there is disagreement, is critical to building fair and equitable communities for everyone, including adequately protecting the needs and interests of vulnerable minority groups. Here, we begin by examining where there is the greatest disagreement on values (Fig.~\ref{fig:disagreement}) before analyzing how different groups of reddit users disagree with others.

Informed by previous work on vulnerable members of online communities~\cite{Lenhart2016OnlineHD, mahar_2018_squadbox}, we hypothesize that Safety will be especially disagreed over, as members who have personally felt unsafe online will perceive the current state of Safety as worse than others, and will rate Safety as more important and more urgent to change. We further hypothesize that newer and less popular community members will generally perceive their communities more negatively than older members, as previous work has found that incorporating newcomers into communities is a significant challenge~\cite{kiene_2016_eternal_september, robert_e_kraut_building_2012}.

\xhdr{Method}
We measure disagreement by computing each response's difference from mean response for the corresponding community. We characterize overall disagreement by averaging across the absolute value of this deviation (MAD). We then further break down which types of community members tend to disagree in which direction by grouping users into approximate quartiles based on their seniority and popularity (with bin edges selected for interpretability), and computing the average deviation from the community mean amongst those groups. We operationalize members' seniority in the community by calculating the number of years since their account was created, and operationalize popularity as the sum of all upvotes received on their posts (called karma on reddit)~\cite{glenski_2017_consumers_curators}.

\xhdr{Results}
We find that, in general, there is strongest consensus on the current state of the community (average MAD=0.17), with greater disagreement on the desired change (average MAD=0.20) and importance of different values (average MAD=0.22; all values adjusted for scale width to enable comparison). There is 13.3\% (1.97 vs. 1.74) more disagreement over the importance of Safety than the importance of all other values, and 47.4\% (2.67 vs. 1.81) more disagreement over the current state of Safety than all other values (Fig.~\ref{fig:disagreement}b,c). Interestingly, there is relative consensus on the desire to improve Safety (Fig.~\ref{fig:disagreement}c). There is strong consensus on the current state of Size (Fig.~\ref{fig:disagreement}b), while there is relative disagreement over the importance and desired change of Size (Fig.~\ref{fig:disagreement}a,c).

\begin{figure}
    \centering
    %  trim={<left> <lower> <right> <upper>}
    \includegraphics[width=\linewidth, trim={3mm 5mm 9mm 7mm clip}]{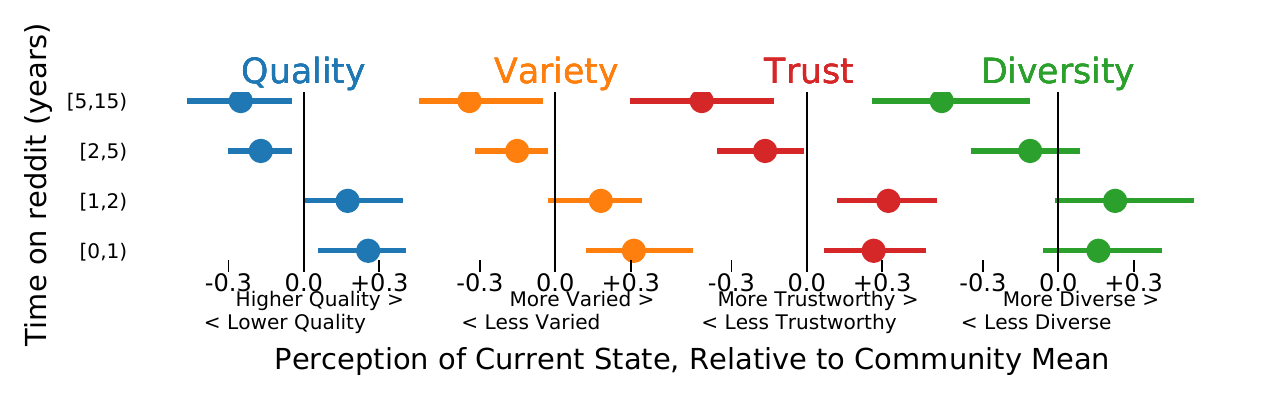}
    \caption{Differences in perception of the current state of communities between new reddit users and those who have been on reddit for longer. Generally, newer reddit users perceive their subs to be 0.55 points higher quality, 0.71 points more varied, 0.74 points more trustworthy, and 0.68 points more diverse compared to older reddit users.}
    \label{fig:user_age}
\end{figure}

When examining differences between senior and junior community members, we demonstrate that junior redditors are generally more positive in their perception of the current state of their communities~(Fig.~\ref{fig:user_age}), in contradiction of our hypothesis that new members' perceptions would be driven by the challenges of assimilation.
Compared to the most senior community members (with at least 5 years of experience), those who joined reddit within the past year perceive their communities to be 0.55 points higher quality, 0.71 points more varied, 0.74 points more trustworthy, and 0.68 points more diverse.
Note that the Current State scale is out of 11 points total, and thus the maximum possible MAD is half the scale width, \ie 5.5. However, the actual distribution of responses is more narrow (see Fig.~\ref{fig:subs_summary}).

\begin{figure}
    \centering
    \includegraphics[width=\linewidth, trim={0 5mm 0 2mm}]{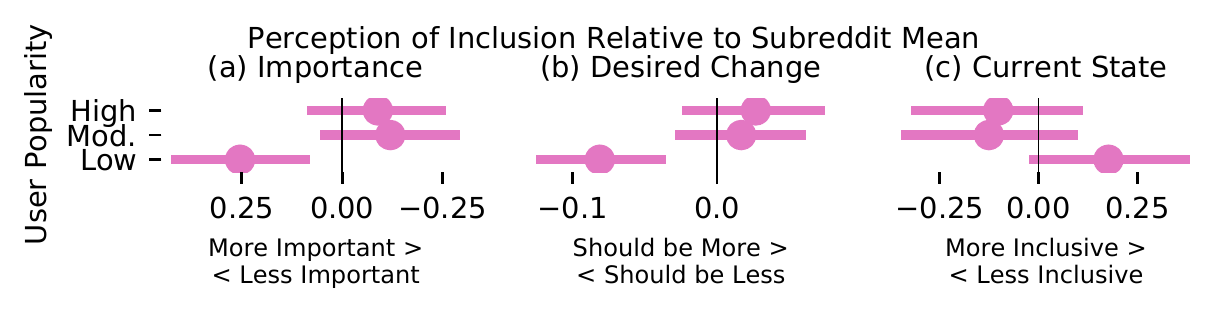}
    \caption{Differences in perceptions of Inclusion across less- and more-popular community members, as measured by account karma, divided into terciles. Relative to more popular users, less popular reddit users perceive Inclusion to be 0.36 points less important (a) and have 0.10 less desire to change Inclusion (b), yet perceive their communities to currently be 0.29 points more inclusive. }
    \label{fig:karma}
\end{figure}

We find significant differences in the perception of Inclusion between less- and more-popular community members (Fig.~\ref{fig:karma}). These differences are especially stark between low-popularity users (in the bottom tercile of karma scores, with less than 67 karma), while differences between moderately and highly popular community members are statistically insignificant. Low-popularity community members place 0.36 points (out of 8) less importance on Inclusion, have 0.10 points (out of 2) less desire for Inclusion to change, and perceive the current state of Inclusion to be 0.29 points (out of 11) better than more popular users.

% moving earlier to get in right spot

\begin{figure*}
    \centering
        %  trim={<left> <lower> <right> <upper>}
    \includegraphics[width=\textwidth, trim={0 5mm 0 2mm}]{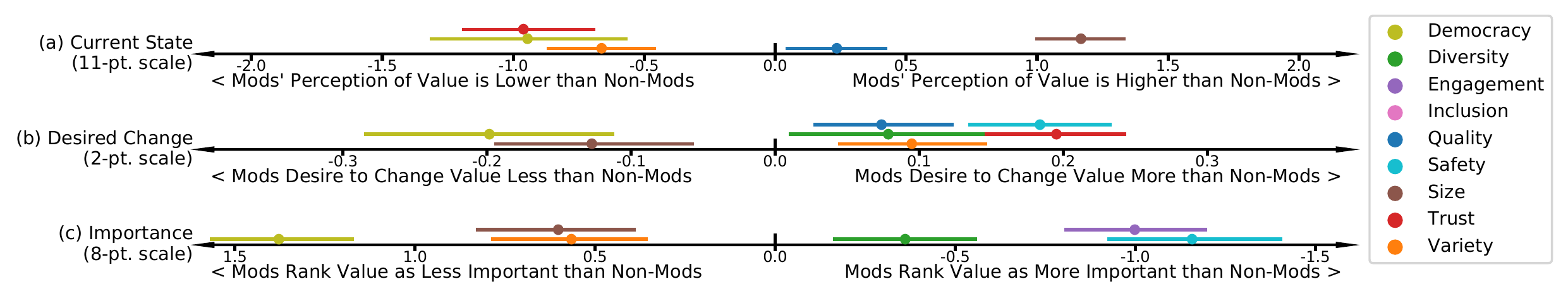}
    \caption{Differences in values between moderators and non-moderators. Moderators believe (a) their communities are 14.5\% less democratic, (b) should be 56.7\% less democratic, and (c) that Democracy is 23.6\% less important, relative to the non-moderator mean in that community. Moderators rank Diversity, Engagement, and Safety as more important to their communities than non-moderator community members (c). Values with CIs overlapping 0 are removed.}
    \label{fig:mods}
\end{figure*}

\xhdr{Implications}
The disagreement over Safety (Fig.~\ref{fig:disagreement}a,b) is a special concern that emphasizes the potential harm of community governance that only responds to the needs of the majority~\cite{Guinier1994TheTO}. While gathering data on past abuse is challenging as well as ethically fraught, it is distinctly probable that the community members most likely to feel that current community Safety is lacking and that Safety ought to be improved are those who have prior negative experiences that made them feel unsafe. While these community members may be a minority, is is critical to design communities that consider and protect their needs. 

Our results also contradict our hypothesis that more senior and more popular users will have a more positive perception of their communities. Instead, we find evidence that it's actually the new reddit users who are most positive in their perception (Fig.~\ref{fig:user_age}), and correspondingly feel that their communities are the most inclusive (Fig.~\ref{fig:karma}c). This is a noteworthy result that suggests that communities on reddit are generally effective in their practices to incorporate new members. However, as we only survey self-identified community members, additional work is needed to reach users who ultimately decided to \textit{not} join a community.

\section{RQ3: How do Moderators Differ in Their Values from Non-moderator Community Members?}\label{sec:moderators}
Volunteer moderators are a key part of any community on reddit, as they bear the primary responsibility of setting rules and enforcing them, a task which frequently brings moderators into conflict with other reddit users~\cite{matias_civic_2019, seering_moderator_2019, seering_2020_self-moderation}. Importantly, moderators also constitute a major part of the governance of communities on reddit~\cite{jhaver_2021_designing}, making their perspective on Democracy especially important. As past work has shown both that much of moderators' interactions with community members are characterized by conflict~\cite{matias_civic_2019} and that affordances for participatory governance are almost entirely absent from reddit~\cite{zhang_policykit_2020}, we hypothesize that moderators will have more negative perceptions of Democracy than non-moderators.

\xhdr{Method}
We identify moderators within our survey responses by scraping users' reddit profile pages, which contain information on the communities each user moderates.
We compute the differences between moderators and non-moderators by grouping responses by community, and, within each community, by taking the difference between all pairs of (mod, non-mod) responses. We compute test statistics and CIs from the resulting set of differences for analysis.

% current         -14.5%  5.57 /  6.51
% change          -56.7%  0.15 /  0.35
% rank            -23.6%  7.22 /  5.84
\xhdr{Results}
We find substantial differences between moderators and non-moderators across all three dimensions of each value: current state, desired change, and importance (Fig.~\ref{fig:mods}). Consistent with our hypothesis, we find that moderators believe their communities \textit{are} 14.5\% (5.57 vs. 6.51) less democratic, \textit{should be} 56.7\% (0.15 vs. 0.35) less democratic, and that Democracy is 23.6\% (7.22 vs. 5.84\footnote{For importance, lower rank values indicate higher importance.}) \textit{less important} than non-moderator members of the same community (relative to the non-moderator mean in that community). When examining all moderator responses (without adjusting for community mean), $2.15 \times$ as many moderators report desiring their communities to be less democratic than non-moderators.
These differences are not limited to Democracy; moderators also more strongly desire to improve the Safety and trustworthiness of their communities than non-moderators (Fig.~\ref{fig:mods}b), and rank Engagement and Safety as more important than non-moderators (Fig.~\ref{fig:mods}c).

\xhdr{Implications}
Moderators are directly able to control many aspects of Democracy in their subreddits (\eg by soliciting community feedback before implementing rule changes), and so their perspective on this value is of special interest. Native tools for enabling formalized community input into governance are lacking from almost all social media platforms, and while some research has attempted to develop such tools~\cite{zhang_policykit_2020}, under current governance paradigms, the adoption of such tools is completely limited by the desire of moderation teams to do so. Moderators also frequently feel overworked~\cite{noauthor_unpaid_nodate, matias_civic_2019} and traumatized by exposure to offensive content~\cite{solon_underpaid_2017}, which may contribute to our findings of a perception of larger community size and lower trustworthiness amongst moderators relative to non-moderators.

\section{RQ4: To What Degree Can Community Values Be Predicted Based on Automatically Measurable Features?}\label{sec:prediction}
Our survey responses (\sect\ref{sec:method}) contain more granular information about community members' values across a far greater set of communities than have been previously collected. However, survey responses are expensive and time consuming to collect and therefore require significant resources to scale. The ability to automatically and accurately predict the importance and desired change of values could be used to inform community design, rule changes, and the implementation of participatory governance practices, while the ability to automatically measure the current state of communities with regards to various values has numerous potential applications, including measuring the impact of interventions.

Throughout this paper, we have demonstrated that communities can vary significantly in their self-reported values, and highlighted general structure in this variation across several potentially generalizable factors.
Here, we investigate how much variation the 14 factors discussed in \sect\ref{sec:values_across_communities}-\ref{sec:moderators}, all of which can be automatically quantified from publicly available data, collectively capture.\footnote{We further experimented with a much larger set of 74 features and found they did not lead to significant performance increases.} A complete list of features used is given in \appendixlink Table~\ref{tab:prediction_features}.

\xhdr{Tasks}
We formulate 27 (importance, current state, and desired change for each of the 9 values) binary classification tasks where the goal is to predict whether a given value is particularly important or unimportant, whether the current state is particularly high or low, and whether the desired change for that value is particularly high or low, for a given subreddit. Each task asks the model to distinguish between the top and bottom quartiles, as for most if not all values, a majority of communities differ only very slightly in their perception of the importance, current state, and desired change of the values. Particularly accurate prediction of small differences is less critical for understanding community values.
As with previous analyses, we aggregate values for each subreddit by averaging across all responses received for that subreddit. To avoid extrapolating from a small number of data points, we filter out communities with fewer than three corresponding survey responses, resulting in 404 communities which are randomly divided with an 80/20 split to create a training and test set. Hyperparameters were chosen through cross-validation on the training data.

\xhdr{Models and Metrics}
We report on an $l_2$-regularized Logistic Regression model with quantile preprocessing, a non-linear quantile transformation which uses the distribution of the training set to spread the data evenly along each feature's axis in the feature space. Missing target values are dropped, and missing features are imputed with the training set mean. Categorical features are one-hot encoded. We also experimented with other models, including neural networks and support vector classifiers, as well as additional preprocessing schemes such as standardization and PCA. We report here on Logistic Regression as overall it performed the best.

\begin{table}[t]
    \small
    \centering
\begin{tabular}{lll|l}
\textbf{Importance} & \textbf{Current State} & \textbf{Desired Change} & \textbf{Overall} \\ \hline
0.660               & 0.673                  & 0.666                   & \textbf{0.667}  
\end{tabular}
    \caption{Quantile-preprocessed Logistic Regression results for the binary classification task on the test set, measured with ROC AUC. Best performance is achieved when predicting the importance of values. In all cases, the model exceeds the performance of a random baseline (0.5 ROC AUC).}
    \label{tab:prediction}
\end{table}

\xhdr{Results}
We find that a Logistic Regression with quantile preprocessing performs the best overall, with an ROC AUC of 0.667 averaged across all tasks. Performance is highest on current state, followed by desired change and then importance (Table~\ref{tab:prediction}). Furthermore, we find that performance is highly variable from value to value; the model is able to accurately predict users' perception of the current Size of the subreddit (ROC AUC 0.936) and the importance of Trust (ROC AUC 0.922), while performance at predicting the importance and current state of Safety is no better than baseline. This is partially due the presence of easily measured proxies for some values (\eg number of contributors is strongly correlated with perception of current Size), while others values, such as Safety, are more nuanced and challenging to automatically measure. A complete table of results for each value is given in \appendixlink ~Table~\ref{tab:pred_extra_results}.

\xhdr{Implications}
These results demonstrate that there is significant structure in how values vary from community to community, and that this structure is predictable using a small number of automatically quantifiable features. Prediction tasks based upon these features could be used to scale research which is informed by the values of the communities it impacts.
However, the overall ROC AUC of $0.667\ll1$ indicates that there is significant remaining structure that is not explained by these features, and further research into what, if any, factors may capture this remaining structure is needed. Features which examine text-based content within subreddits, and graph-based features computed using subreddit membership are two promising avenues for future experimentation. We make our dataset public\footnote{\scriptsize \url{https://behavioral-data.github.io/reddit_values_surveys_public/}} to support further research.
\section{Discussion \& Conclusion}\label{sec:discussion}

\xhdr{Diversity of Communities}
Our study reveals that the set of communities surveyed have remarkably diverse values (Figs.~\ref{fig:subs_summary},\ref{fig:community_categories},\ref{fig:sub_age_size}). This underlines that there is no global set of values common to all online communities; what is important to one may be unimportant or even detrimental to another. Researchers, community leaders, and platforms alike must consider the specific context of the community and its needs before implementing changes. 

\xhdr{Protecting Vulnerable Minorities}
Community members are especially divided on the importance and current state of Safety (Fig.~\ref{fig:disagreement}a,b), with 47.4\% more disagreement over the current state of Safety than any other value in our survey. 
% Although 
Because in many cases community members who feel their communities are unsafe are in the minority, care must be taken to protect the interests of these vulnerable groups. Although additional research is needed on this important topic, some work has shown that even simple interventions such as automated welcome messages can help support minority groups~\cite{matias_study_2020}.

\xhdr{Participatory Governance}
Volunteer moderators play an important role in community governance on reddit~\cite{matias_civic_2019}.
On the other hand, both formal and informal opportunities for non-moderators to influence decision making in their communities are quite rare~\cite{zhang_policykit_2020}. We find that in general, while non-moderators desire to have more Democracy in their communities, moderators are 56.7\% less in favor of increased Democracy (Fig.~\ref{fig:mods}b). This discrepancy could pose a challenge to increasing participatory governance; more research is needed on why moderators are less approving of Democracy, and what changes are needed to mitigate this difference. 

\xhdr{Limitations}
Our research is carried out only on reddit; additional work is needed to understand how our findings generalize to other platforms such as Twitter and Facebook. While we made every effort to recruit a diverse set of participants by using multiple recruiting methods, our work is still subject to some potential bias from groups of people who were not included in our study. One source of this bias is the limitations in who we can target ads to, as reddit restricts advertising to members of communities focused on porn and other controversial topics. We also recognize that in our analyses, when we filter out communities with fewer responses, we're disproportionately excluding smaller communities, however this filtering step is necessary to reliably assess consensus and disagreement (\sect\ref{sec:instrument}).

\xhdr{Conclusion}
Online communities are extraordinarily varied, and the importance they place on different values reflects this variety. As such, what is good for one community may be harmful to another. In this work, we surveyed 2,796 reddit users to characterize how their values vary within and across 2,151 different communities. By combining these survey responses with publicly available reddit data, we examined differences between communities focused different topics, and measured where there is consensus and disagreement over different values. We compared moderators' values to non-moderators' values, identifying challenges for the implementation of participatory governance online. We make our dataset public to support future research.

\section*{Acknowledgements}
This research was supported in part by the Office for Naval Research (\#N00014-21-1-2154), NSF grant IIS-1901386, NSF grant CNS-2025022, the Bill \& Melinda Gates Foundation (INV-004841), a Microsoft AI for Accessibility grant, and a Garvey Institute Innovation grant.
 % todo uncomment for camera ready

%%
{\footnotesize
\bibliography{bibliography}
}

\clearpage

\appendix
\onecolumn

\section{Categorizing Subreddit Topics}\label{app:categories}
To operationalize higher-level notions of community topic and focus,\footnote{We additionally experimented with pre-computed subreddit embeddings~\cite{kumar_community_2018, martin_community2vec_2017, waller_generalists_2019}. We found these representations were unable to differentiate between communities based on their values.} we manually investigated each of the 122 subreddits for which we received responses from at least 10 different community members. Among the author team, we iteratively clustered these communities until there was agreement on 6 different and mutually exclusive categories. We were unable to come up with other categories that were relevant to a significant fraction of these communities.

% Hobby         53
% Discussion    18
% Pictures      17
% News          15
% Memes         11
% Identity       8

\begin{itemize}
    \item \textbf{Hobby} Communities for people interested in specific games and hobbies (53 communities, \eg~/r/nba, /r/bicycling)
    \item \textbf{Discussion} Communities which focus on question answering and discussion (18 communities, \eg~/r/AskReddit, /r/re\-lationship\_advice)
    \item \textbf{Media-sharing} Communities for posting pictures and video of different things (17 communities, \eg~/r/pics, /r/Crappy\-Design)
    \item \textbf{News} Communities which share news, research, and data (15 communities, \eg~/r/worldnews, /r/science)
    \item \textbf{Meme} Communities which are primarily for memes and shitposting (11 communities, \eg~/r/dankmemes, /r/me\_irl)
    \item \textbf{Identity-based} Communities which are primarily for specific groups of people (8 communities, \eg~/r/india, /r/teen\-agers)
\end{itemize}

\section{Power Analysis to Determine Validity of Responses}\label{app:num_responses}

\begin{figure*}[!h]
    \centering
    \includegraphics[width=.6\textwidth]{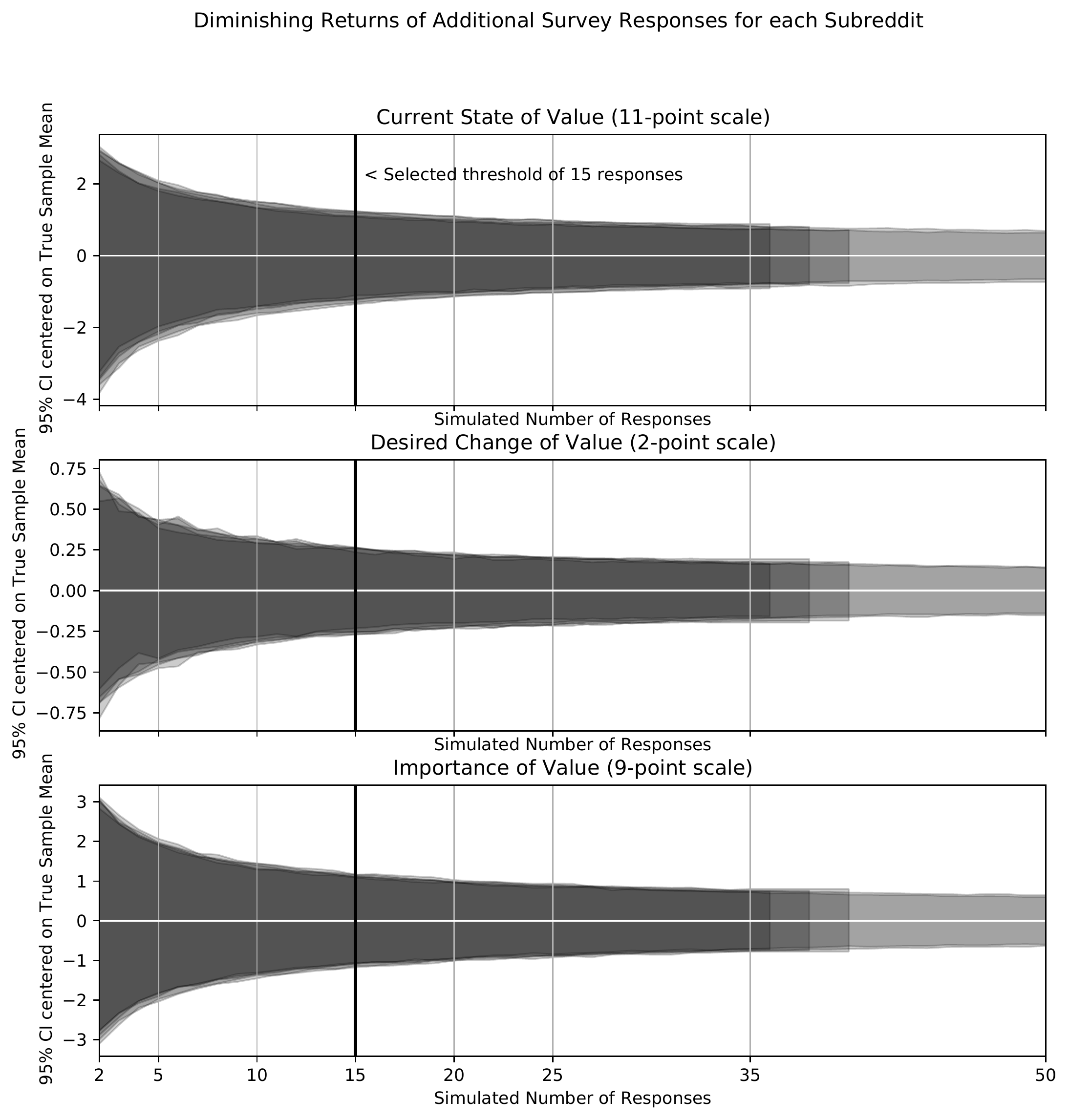}
    \caption{Using responses from the 5 subreddits with more than 35 responses each, we randomly downsampled (1,000-fold bootstrapping) responses to estimate the sample variance when collecting fewer responses. We found that beyond 15 responses per subreddit, sample variance does not decrease significantly, and so we select this threshold for our analyses.}
    \label{fig:num_responses}
\end{figure*}

\clearpage
\section{Participant Recruiting and Incentives}\label{app:recruiting}

\begin{figure}[h]
     \centering
     \begin{subfigure}[b]{0.48\textwidth}
         \centering
         \includegraphics[width=.95\textwidth]{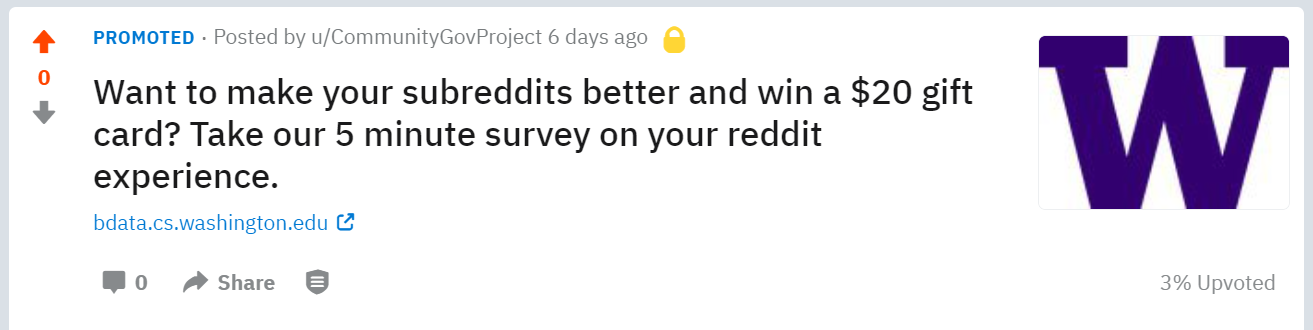}
         %\caption{$y=x$}
         \label{fig:ad_1}
     \end{subfigure}
     \hfill
     \begin{subfigure}[b]{0.48\textwidth}
         \centering
         \includegraphics[width=.95\textwidth]{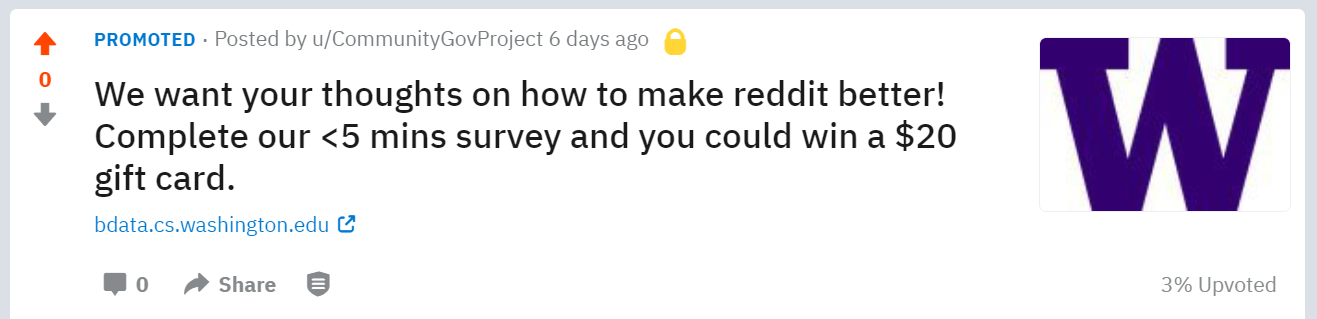}
         %\caption{$y=3sinx$}
         \label{fig:ad_2}
     \end{subfigure}
     \caption{Reddit advertisements used to recruit participants.}
     \label{fig:recruitment}
\end{figure}

Survey respondents were recruited primarily through reddit advertisements and private messages (PMs), which are displayed to reddit users on both the website as well as the reddit mobile app. We used several different titles for the ads, Appendix Fig.~\ref{fig:recruitment} shows examples. We ran three different recruitment campaigns: (1) a general campaign targeted at all reddit users and designed to capture responses from members of a wide range of subreddits, (2) a specific campaign intended to increase the number of responses received for the most popular subreddits, conducted by creating separate ads for each of the 300 largest subreddits, and (3) a moderator recruitment campaign to encourage participation specifically from community moderators, who were recruited via PMs sent to each of the 100 largest subreddits).

Survey responses were gathered from May-July 2021. In total, 2,769 people participated, with the participants answering questions for 2.15 subreddits on average, for a total of 5,962 subreddit-responses across 2,151 unique subreddits. 562 responses (20.30\%) were recruited via the general campaign, 2,022 (73.02\%) were recruited the specific campaign, 81 (2.93\%) were recruited via the moderator campaign, and 104 (3.80\%) were recruited via friend referrals and word of mouth. The median completion time was approximately 8 minutes. 97.33\% of respondents provided their username and consented to the inclusion of their post and comment history in our research.

To incentivize participation, we raffled off a \$100 Amazon gift card and 5$\times$ \$20 Amazon gift cards to participants who completed the survey. Participants were offered additional raffle tickets for recruiting their friends to participate as well. Winners were contacted via reddit PM, and those outside of the US were offered a gift card of equivalent value in their local currency. 

\clearpage
\section{Details on Prediction Tasks}

\begin{table}[h]
    \small
    \centering
\begin{tabular}{ll|r}
Value & Dimension &                                         ROC AUC \\
\hline
Democracy & Current State &                                    0.622 \\
        & Desired Change &                                    0.684 \\
        & Importance &                                    0.541 \\
Diversity & Current State &                                    0.634 \\
        & Desired Change &                                    0.800 \\
        & Importance &                                    0.716 \\
Engagement & Current State &                                    0.635 \\
        & Desired Change &                                    0.532 \\
        & Importance &                                    0.642 \\
Inclusion & Current State &                                    0.730 \\
        & Desired Change &                                    0.708 \\
        & Importance &                                    0.555 \\
Quality & Current State &                                    0.725 \\
        & Desired Change &                                    0.624 \\
        & Importance &                                    0.677 \\
Safety & Current State &                                    0.441 \\
        & Desired Change &                                    0.714 \\
        & Importance &                                    0.391 \\
Size & Current State &                                    0.936 \\
        & Desired Change &                                    0.655 \\
        & Importance &                                    0.661 \\
Trust & Current State &                                    0.709 \\
        & Desired Change &                                    0.688 \\
        & Importance &                                    0.922 \\
Variety & Current State &                                    0.625 \\
        & Desired Change &                                    0.589 \\
        & Importance &                                    0.838 \\
\hline
\end{tabular}
    \caption{Task-level results (ROC AUC) for the Logistic Regression model on our 27 prediction tasks. }
    \label{tab:pred_extra_results}
\end{table}

\begin{table*}
    \small
    \centering
    \begin{tabular}{r|l}
\texttt{sub\_num\_posts}               & The number of posts in the subreddit. \\
\texttt{sub\_num\_removed\_posts}      & The number of posts removed by a moderator in the subreddit. \\
\texttt{sub\_num\_deleted\_posts}      & The number of posts deleted by their author in the subreddit. \\
\texttt{sub\_num\_selfposts}           & The number of selfposts (text-posts) in the subreddit. \\
\texttt{sub\_num\_linkposts}           & The number of posts which link to external websites. \\
\texttt{sub\_num\_comments}            & The number of comments in the subreddit. \\
\texttt{sub\_num\_removed\_comments}   & The number of comments removed by a moderator. \\
\texttt{sub\_num\_deleted\_comments}   & The number of comments deleted by their author. \\
\texttt{sub\_distinct\_users}          & The number of distinct contributors to the subreddit. \\
\texttt{sub\_num\_subscribers}         & The number of users who `subscribe' to the subreddit. \\
\texttt{sub\_age}                      & The number of days since the subreddit was founded. \\
\texttt{sub\_topic\_specificity}       & The manually-categorized specificity of the topic of the subreddit, on an 3-point scale. \\
\texttt{sub\_topic\_category}          & The manually-categorized (see \sect\ref{sec:user_sub_feats}) topic of the subreddit. \\
    \end{tabular}
    \caption{Descriptions of features used in the prediction tasks (\sect\ref{sec:prediction}).}
    \label{tab:prediction_features}
\end{table*}

\end{document}